\documentclass[pre,aps,showkeys,showpacs,floatfix]{revtex4}

\usepackage{graphicx}
\usepackage[latin1]{inputenc}
\usepackage{amsmath}
\usepackage{amsfonts}
\usepackage{amssymb}
\usepackage{epsf}
\usepackage{epsfig}

\newcommand{\be}{\begin{equation}}
\newcommand{\ee}{\end{equation}}
\newcommand{\ba}{\begin{eqnarray}}
\newcommand{\ea}{\end{eqnarray}}

\usepackage{soul}

\begin{document}

\title{Application of Fractional Calculus to Reaction-Subdiffusion Processes and Morphogen Gradient Formation}
\author{S. B. Yuste$^{1}$, E. Abad$^{1}$, and Katja
Lindenberg$^{2}$} \affiliation{$^{(1)}$ Departamento de F\'{\i}sica, Universidad de Extremadura, E-06071 Badajoz, Spain\\
$^{(2)}$Department of Chemistry and Biochemistry, and BioCircuits Institute, University of California San Diego, 9500 Gilman
Drive, La Jolla, CA 92093-0340, USA}

\date{\today}

\begin{abstract}                
It is a well known fact that subdiffusion equations in terms of fractional derivatives can be obtained from Continuous Time
Random Walk (CTRW) models with long-tailed waiting time distributions. Over the last years various authors have shown that
extensions of such CTRW models incorporating reactive processes to the mesoscopic transport equations may lead to
non-intuitive reaction-subdiffusion equations. In particular, one such equation has been recently derived for a subdiffusive
random walker subject to a linear (first-order) death process. We take this equation as a starting point to study the
developmental biology key problem of morphogen gradient formation, both for the uniform case where the morphogen degradation
rate coefficient (reactivity) is constant and for the non-uniform case (position-dependent reactivity). In the uniform case we
obtain exponentially decreasing stationary concentration profiles and we study their robustness with respect to perturbations
in the incoming morphogen flux. In the non-uniform case we find a rich phenomenology at the level of the stationary profiles.
We conclude that the analytic form of the long-time morphogen concentration profiles is very sensitive to the spatial
dependence of the reactivity and the specific value of the anomalous diffusion coefficient.

\end{abstract}



\maketitle

\section{Introduction}

Fractional diffusion equations are a powerful tool to study anomalous transport processes, i.e., processes in which the mean
square displacement $\langle x^2 \rangle$ of a randomly moving particle displays the long time-behavior $\langle x^2 \rangle
\sim K_\gamma t^\gamma$, where $\gamma$ is the anomalous diffusion exponent and $K_\gamma$ is the so-called anomalous
diffusion coefficient. When $0<\gamma<1$, one has sublinear growth of $\langle x^2 \rangle$ (subdiffusion), while for
$\gamma>1$ one speaks of superdiffusion. As it is well known, the classical diffusion equation corresponding to the $\gamma=1$
case can be obtained from an average over the trajectories of a Markovian random walk in the limit of large time scales and
long displacements. In contrast, stochastic transport processes governed by anomalous diffusion equations reflect memory
effects at the microscopic level. In particular, one can show that a suitably defined non-Markovian hopping process, namely a
Continuous Time Random Walk (CTRW) with a long-tailed waiting time distribution, yields a subdiffusion equation in terms of
the Riemann-Liouville fractional derivative \cite{MetzlerKlafterPhysReport}. This fractional subdiffusion equation can be
taken as a starting point to deal with a number of biologically relevant problems, e.g. the localization of a target protein
by a sea of subdiffusively moving ligands in the intracellular environment \cite{targetyuste,target09}. In this case, the
complexity of the cell medium results in the ligands encountering a large number of obstacles, barriers, etc. in the course of
their trajectories. In the framework of CTRW models, the effect of this crowded environment can be partly captured using
waiting time distributions; subsequent averaging of the resulting equations over trajectories then leads to the associated
fractional subdiffusion equations.

While anomalous diffusion and in particular subdiffusive processes play a central role in Nature as a manifestation of
underlying memory effects at a microscopic level, the situation where the particles simultaneously undergo anomalous transport
{\it and} reaction (understood as a particle creation, destruction or transformation process) is also very common and
important from the point of view of biological applications. In the example of the target protein and the ligands given above,
one could allow e.g. for the possibility of the ligands undergoing a degradation process as they sweep the cell medium.
Degradation implies a change in chemical structure which results in the ligands losing their ability to interact with the
target; for practical purposes this kind of transformation can therefore be regarded as a ``death" or ``evanescence" process.

In what follows, we shall focus on yet another biological process where degradation/death plays a central role, namely
morphogen gradient formation. The location, differentiation and fate of many embryonic cells is governed by the spatial
distribution of special signaling molecules called morphogens. Standard models of morphogen gradient formation assume that a
specific part of the embryo secrets morphogens at a constant rate. The secreted morphogens then undergo degradation as they
disseminate through the tissue and a concentration gradient builds up. Different target genes in the embryonic cells are
activated above different morphogen concentration thresholds, implying that the cell response to the local environment will
depend on how large the concentration is. Thanks to this differential response, cells are able to interpret the morphogen
gradient and translate it into specific ``code" for their further development via the expression of the relevant genes.

Traditional models of morphogen gradient formation are based on classical diffusion equations with a linear degradation term.
Here, we aim to go one step further and allow for the possibility of anomalous transport, as memory effects are likely to
strongly influence the stochastic motion of morphogens in the complex embryonic environment. A remarkable property of
morphogen gradients is their robustness against changes or fluctuations in the rate of morphogen production or degradation. An
interesting question which we aim to study is the interplay between subdiffusion and robustness with respect to such
perturbations.

If one accepts the idea that morphogens perform subdiffusive motion as a result of strong dispersion in their waiting times
between consecutive jumps, great caution must be exercised when incorporating the effect of a simultaneous degradation process
to the transport equations because of the non-Markovian character of the latter. Several recent works indeed illustrate that
heuristic equations where one has separate terms for the reaction and the transport process may lead to unphysical results,
e.g. negative particle concentrations (see e.g. \cite{Langlands}). Therefore, the derivation of physically correct (but not
necessarily intuitive) reaction-subdiffusion equations calls for the use of an extended CTRW formalism where the effect of
reaction is incorporated at a {\it mesoscopic} level of description. In a recent work \cite{Nuestropaper} the authors have
shown by means of Fourier-Laplace techniques that CTRW models extended in such a way yield equations which (in addition to a
standard, purely reactive term) display a mixed reaction-transport term containing both the reaction rate coefficient
(reactivity) and a Riemann-Liouville fractional derivative with respect to time.

The remainder of the paper is organized as follows. We first give a brief reminder of classical reaction diffusion equations
used for modeling of morphogen gradients and subsequently discuss how to extend such equations to account for anomalous
transport via fractional derivatives. We subsequently focus on the specific case of uniform reactivity and assess the
robustness of the resulting stationary concentration profiles with respect to perturbations of the incoming morphogen flux.
Next, we turn to the non-uniform case and discuss the long-time behaviour of the profiles for several specific situations,
namely the case of a piecewise constant reactivity (which not always leads to a stationary profile) and the case of a decaying
reactivity respectively given by an exponential and a power law. In some cases, we also provide numerical simulation results
based on a CTRW model for evanescent particles and find excellent agreement with the analytic results obtained from the
fractional diffusion equation approach. Finally, a summary of results and possible avenues for future research in this area
are given in the conclusions section.

\section{Classical reaction-diffusion equation with linear degradation}

The cornerstone of many studies concerning morphogen gradients is the classical one-dimensional reaction-diffusion equation

\be \label{classrdeq} \frac{\partial c(x,t) }{\partial t} = K_1 \frac{\partial^2 c(x,t)}{\partial x^2}-k(x,t) \, c(x,t), \ee

where the evolution of the concentration $c(x,t)$ is described by a Fickian term (characterized by a classical diffusion
coefficient $K_1$) and a linear degradation term (characterized by the reactivity $k(x,t)$). Eq. (\ref{classrdeq}) is then
solved subject to the radiation-type boundary condition

\be \label{radbc} -K_1\left.\frac{\partial c(x,t)}{\partial x}\right|_{x=0+}+ K_1\left.\frac{\partial c(x,t)}{\partial
x}\right|_{x=0-}=j_0. \ee

This boundary condition simply states that a constant flux of morphogens $j_0$ is injected at the origin $x=0$. The simplest
case is given by a constant degradation rate $k(x,t)=k$, which yields the exponentially decaying stationary profiles:

\be \label{stexppr} c(x,\infty)= \frac{j_0}{2\sqrt{k K_1}} \; e^{-\sqrt{\frac{k}{K_1}} |x| }. \ee

Despite its simplicity, the exponential dependence of Eq. (\ref{stexppr}) captures surprisingly well the rapid concentration
decay displayed by real profiles. However, the separate determination of $K_1$ and $k$ poses significant experimental
difficulties and in most cases only the characteristic length $\sqrt{K_1/k}$ can be unambiguously determined. This opens the
door to the possibility of considering non-Markovian generalizations of Eq. (\ref{classrdeq}) which are also compatible with
the experimental results for the characteristic length.

\section{ Fractional reaction-subdiffusion equation with uniform reactivity}

\subsection{Derivation from a mesoscopic CTRW model with reaction}

The starting point to derive the reaction-subdiffusion equation is the fundamental equation of a particle performing a
continuous time random walk in the presence of a first order evanescence reaction (degradation). Consider the stochastic
motion of an evanescent particle performing nearest-neighbor jumps on an infinite one-dimensional lattice. We shall hereafter
denote the distribution of its waiting time between consecutive jumps by $\psi(t)$.The probability that a particle starting at
site $0$ arrives at site $i$ in the time interval between $t$ and $t+dt$ can be written as $q(i,t)\,dt$, where the arrival
density $q$ obeys the following integral difference equation:

\be \label{inteq1} q(i,t)=\frac{1}{2}\sum_{j=0}^1\int_0^t dt' \psi_{RW}(t-t')q(i+(-1)^j,t')+\delta_{i,0}. \ee

In this equation, $\psi_{RW}(t)=\psi(t)e^{-k\,t}$ is the probability per unit time that a particle found at a given site at
time $t=0$ has performed a jump up to time $t$ in the presence of a uniform evanescence reaction (since evanescence and jump
are independent processes, the probability of jump is simply multiplied by the survival probability $e^{-k\,t}$).  Taking the
diffusive limit of Eq. (\ref{inteq1}) one can show via suitable Fourier-Laplace techniques \cite{Langlands} that the
probability $w(x,t|0,0)$ to find the particle at position $x$ after a time $t$ given that it was initially at $x=0$ obeys the
following equation:

\ba \frac{\partial w(x,t|0,0)} {\partial t}&& =K_\gamma e^{-k t}   ~_{0}{\cal D}_t^{1-\gamma} e^{k t} \frac{\partial^2
}{\partial x^2} w(x,t|0,0) \nonumber \\ \label{unifreacsubdeq0} &&-k  w(x,t|0,0). \ea

The operator $~_{0}\,{\cal D}_t^{1-\gamma}$ is defined via the equation

\be {\cal L}_{u\to t}^{-1}\left\{u^{1-\gamma}\tilde{y}(u)\right\}=~_{0}\,{\cal D}_t^{1-\gamma} y(t), \ee

where $\tilde{y}(u)$ is the Laplace transform of the function $y(t)$ and ${\cal L}_{u\to t}^{-1}\left\{\cdot\right\}$ denotes
the inverse Laplace transform. The operator $~_{0}\,{\cal D}_t^{1-\gamma}$ is closely related to the Riemann-Liouville
fractional derivative

\be ~_{0}D_t^{1-\gamma} f(x,t) =\frac{1}{\Gamma(\gamma)}\frac{\partial}{\partial t} \int_0^t dt'\,
\frac{f(x,t')}{(t-t')^{1-\gamma}}. \ee

In fact, $~_{0}{\cal D}_t^{1-\gamma}$ and $_{0}D_t^{1-\gamma}$ are the same when applied to sufficiently regular functions
$f(t)$, as determined by the condition $\lim_{t\to 0} \int_0^t dt' (t-t')^{\gamma-1} f(t')=0$. This condition is actually
fulfilled by all functions of $t$ relevant to the morphogen problem, hence we shall use $_{0}D_t^{1-\gamma}$ in place of
$~_{0}{\cal D}_t^{1-\gamma}$ in what follows.

If one is dealing with more than one particle, the concentration $c(x,t)$ follows the same kinetics as above, i.e. \cite{SSS,
Langlands}

\be \label{unifreacsubdeq} \frac{\partial c(x,t)} {\partial t}=K_\gamma e^{-k t}   ~_{0}D_t^{1-\gamma} e^{k t}
\frac{\partial^2 }{\partial x^2} c(x,t)-k  c(x,t). \ee

As one can see, this equation is a non-trivial extension of Eq. (\ref{classrdeq}) for the case of anomalous subdiffusion with
constant reactivity $k$. In the normal diffusion limit $\gamma \to 1$ the Riemann-Liouville operator reduces to unit and one
recovers Eq. (\ref{classrdeq}) with a constant $k$. On the other hand, in the absence of reaction ($k\to 0$) Eq.
(\ref{unifreacsubdeq}) reduces to the standard fractional diffusion equation, which yields sublinear growth of $\langle x^2
\rangle$.

Turning now to the morphogen problem, Eq. (\ref{unifreacsubdeq}) is to be solved subject to the boundary condition
(\ref{radbc}). The solution $c(x,t)$ for the case of a particle source can be obtained from the propagator solution
$c_P(x,t)\equiv G(x,t)$ (corresponding the initial condition $G(x,0)=\delta(x)$) via the relation $\tilde c(x,u)=j_0\,\tilde
G(x,u)/u$ between the Laplace transforms. The solution in Laplace space is found to be

\be \label{cxukcte} \tilde c(x,u)=  \frac{j_0}{2}
\frac{(u+k)^{\gamma/2-1}}{u\sqrt{K_\gamma}}\exp\left[-(u+k)^{\gamma/2}/\sqrt{K_\gamma}|x|\right]. \ee

The stationary solution is obtained from the final value theorem for the Laplace transform:

\be \label{stprofile} c_s(x)=\lim_{u\to 0} u \tilde c(x,u)= \frac{j_0}{2} \frac{k^{\gamma/2-1}}{\sqrt{K_\gamma}}\exp\left[-|x|
k^{\gamma/2}/\sqrt{K_\gamma}\right]. \ee

Eq. (\ref{stprofile}) generalizes the exponential profile described by Eq. (\ref{stexppr}) . Steady state profiles are thus
seen to also exist in the presence of anomalous diffusion, as opposed to what had been suggested in some previous works
\cite{HBB}.

\subsection{Robustness of stationary profiles}

Using Eq. (\ref{stprofile}) it is possible to study the robustness of the concentration profiles with respect to a
perturbation in the incoming flux $j_0$. To this end, we take a reference value $c_\times$ of the concentration and assess how
large the shift of the associated position

\be \label{assxval} x=\sqrt{\frac{K_\gamma}{k^\gamma}}\ln\left(\frac{k^{\gamma/2-1}j_0} {2c_\times\,\sqrt{K_\gamma}}\right)
\ee

becomes when $j_0$ is perturbed; the larger the shift, the smaller the robustness of the profile. The latter can thus be
characterized by the inverse of the relative change of $x$ with respect to a characteristic length $a$ of the problem (e.g.
the linear size of a cell), i.e.

\be {\cal R}_{j_0}= a \left(j_0\frac{\partial x}{\partial j_0}\right)^{-1} \ee

Inserting Eq. (\ref{assxval}) into this definition we find

\be {\cal R}_{j_0} \propto \sqrt{\frac{k^\gamma}{K_\gamma}}. \ee

\section{Fractional reaction-subdiffusion equation with non-uniform reactivity}

Seki et al. \cite{SekiJPCM07} have shown that a CTRW process described by a generalization of Eq. (\ref{inteq1}), namely

\ba q(i,t)&=& \frac{1}{2}\sum_{j=0}^1\int_0^t dt' \psi_{RW}(i+(-1)^j, t-t')q(i+(-1)^j,t')\nonumber \\ \label{inteq2} &&
+\delta_{i,0} \ea

with $\psi_{RW}(i,t)=\psi(t)e^{-k(i)\,t}$ yields the following reaction-subdiffusion equation:

\ba \frac{\partial c(x,t)}{\partial t} &&= K_\gamma \frac{\partial^2 }{\partial x^2}\left[e^{-k(x) t}
~_{0}D_t^{1-\gamma}\left( e^{k(x) t}  c(x,t)\right) \right] \nonumber \\ \label{sekiEq} &&-k(x)  c(x,t). \ea

In order to tackle the corresponding morphogen problem, it is first necessary to find the propagator solution of Eq.
(\ref{sekiEq}). To this end it is convenient to introduce a new function $v(x,t)$ defined via the transformation

\be \label{vxudef} \tilde v(x,u)=[u+k(x)]^{1-\gamma} \tilde c(x,u) \ee

in Laplace space. This function is readily found to fulfil the equation

\be \label{vxunudelta} [u+k(x)]^{\gamma} \tilde v(x,u)-\delta(x)= K_\gamma  \frac{\partial^2 }{\partial x^2}  \tilde v(x,u).
\ee

In what follows, Eq. (\ref{vxunudelta}) will be used to investigate the effect of a non-uniform reactivity for several special
cases.

\subsection{Piecewise constant reactivity}

Here, we assume that the reactivity is given by a superposition of Heaviside functions, i.e. $k(x)=k_0 H(R-|x|)+k_1 H(|x|-R)$.
In region $0$ ($0\le |x|<R$) one has $k(x)=k_0>0$, whereas  in region $1$ ($|x|\ge R$) one has $k(x)=k_1 \ge 0$. Let us
respectively denote by $\tilde v_0(x,u)$ and $\tilde v_1(x,u)$ the solutions of Eq. (\ref{vxunudelta}) in the regions $0$ and
$1$. These functions must fulfil the continuity conditions

\be \label{contcond} \tilde v_0(R,u)= \tilde v_1(R,u) \ee

and

\be \left .\frac{\partial \tilde v_0(x,u)}{\partial x}\right|_{|x|=R} =  \left .\frac{\partial \tilde v_1(x,u)}{\partial x}
\right|_{|x|=R}. \ee

In contrast, an integration of Eq. (\ref{vxunudelta}) across the origin shows that the solution must be discontinuous there:

\be \label{disccond} \left.\frac{\partial \tilde v_0(x,u)}{\partial x}\right|_{x=0+}- \left.\frac{\partial \tilde
v_0(x,u)}{\partial x}\right|_{x=0-}=-\frac{1}{K_\gamma}. \ee

Using Eqs. (\ref{contcond})-(\ref{disccond}) one can find explicit expressions for the Laplace transforms $\tilde v(x,u)$,
$\tilde G(x,u)$ and $\tilde c(x,u)=j_0 \tilde G(x,u)/u$. For $|x|<R$ one gets the stationary biexponential solution

\be \label{relcsv0} c_s(x)=  j_0\;k_0^{\gamma-1} \tilde v_0(x,u\to 0) \ee

with

\be \label{biexpeq} \tilde v_0(x,u)=A_0 e^{-\alpha_0 x}+B_0 e^{\alpha_0 x},\qquad \alpha_0^2=\frac{(u+k_0)^\gamma}{K_\gamma}
\ee

The characteristic constants are

\ba A_0&& = \frac{\left(k_0+u\right){}^{-\gamma /2}/2 \sqrt{K_{\gamma }}} { \left(\frac{2}{ -1- \left(k_1+u\right){}^{\gamma
/2} \left(k_0+u\right){}^{-\gamma /2}} +1\right) e^{-\frac{2 R \left(k_0+u\right){}^{\gamma /2}}{\sqrt{K_{\gamma }}}}\!\!+1
\!\!} \nonumber \\ B_0&&= \frac{\left(k_0+u\right){}^{-\gamma /2}/2 \sqrt{K_{\gamma }}}{ \left(\frac{2}{1-
\left(k_0+u\right){}^{-\gamma /2} \left(k_1+u\right){}^{\gamma /2}}-1\right) e^{\frac{2 R \left(k_0+u\right){}^{\gamma
/2}}{\sqrt{K_{\gamma }}}}-1 } \nonumber \ea

For $|x|>R$ we shall distinguish two subcases with different physical behaviour. For $k_1>0$ one asymptotically gets the
exponential decay law

\be c_s(x)=c(x,t\to\infty) \propto   e^{-k_1^{\gamma/2} (x-R)/\sqrt{K_\gamma}}. \ee

In contrast, for $k_1=0$ the behaviour is different. For normal diffusion the profile becomes constant for large $|x|$, i.e.,

\be c_s(x) \propto  j_0 \qquad (k_1=0,\gamma=1,|x|\ge R). \ee

\begin{figure}
\begin{center}
\includegraphics[width=0.5\textwidth]{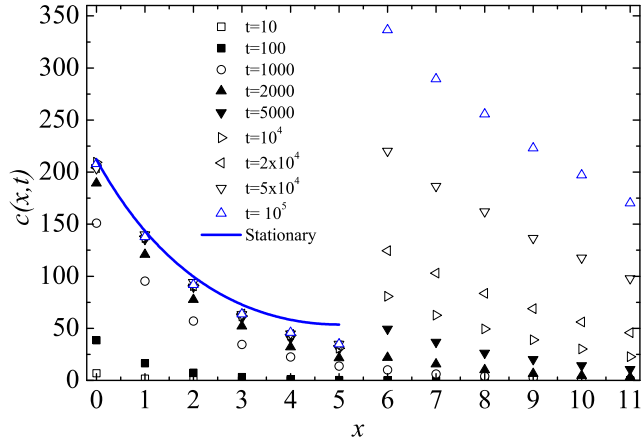}
\end{center}
\caption{Simulation results (symbols) of $c(x,t)$ for a step reactivity
  [$k(x)=k_0 H(R-x)$] with  $k_0=1/1000$ and $R=5$ for $\gamma=1/2$ (only
  values for x>0 are shown). The particles are simulated by means of a CTRW model with the Pareto waiting time distribution
  $\psi(t)=\gamma/(1+t)^{1+\gamma}$ and equiprobable jumps $\{-1,0,1\}$. These parameters lead to the $K_\gamma$-value
  $1/\sqrt{9\pi}$. The solid line corresponds to the theoretical prediction for the steady-state profile when $x<R$. For $x>R$
  no stationary profile is developed.  The convergence of the simulation results to the stationary profile for $x<R$ is very
  slow for values of $x$ close to the discontinuity at $x=R$. No adjustable parameters were used.}
\label{fig:Escalon1} \end{figure}

However, when the diffusion is anomalous one has

\be \label{nostt} c(x,t\to\infty) \propto  j_0 \; t^{1-\gamma}  \qquad (k_1=0, \gamma<1, |x|\ge R), \ee

i.e., there is no steady state! In view of Eq. (\ref{relcsv0}) and (\ref{nostt}), we conclude that the profile is
discontinuous at $x=R$. This behaviour is confirmed by numerical simulations (see Fig. \ref{fig:Escalon1}).

\subsection{Exponentially decaying reactivity}

Here, we assume the decay law $k(x)=k_0\,e^{-\beta |x|}$. While in this case Eq. (\ref{vxunudelta}) does not seem exactly
solvable for finite $u$, it is possible to find an exact expression of the steady state profile by techniques similar to the
ones used above. The final result is

\be \label{bessprof} c_s(x)=j_0 \frac{k_0^{\gamma/2-1}}{2K_\gamma^{1/2}} \frac{I_0\left(\alpha k_0^{\gamma/2} e^{-\beta\gamma
|x|/2}\right)} {I_1\left(\alpha k_0^{\gamma/2}\right)} \; e^{-(\gamma-1) \beta |x|}, \ee

where the $I_n$'s are modified Bessel functions and $\alpha=2/(\beta \gamma \sqrt{K_\gamma})$. As in the case of piecewise
reactivity with $k_1=0$, this expression displays a different behaviour for normal and anomalous diffusion. In the normal
diffusion case ($\gamma=1$) one gets a monotonically decreasing profile from the concentration value

\be c_s(x=0)= \frac{j_0 }{\sqrt{4 k_0 K_\gamma }} \frac{I_0\left(\alpha k_0^{1/2}\right)} {I_1\left(\alpha k_0^{1/2}\right)}
\ee

at the origin to the limiting value

\be c_s(x\to\pm \infty)=\frac{j_0 }{\sqrt{4 k_0 K_\gamma }} \frac{1}{I_1\left(\alpha k_0^{1/2}\right)}. \ee

\begin{figure}[ht]
\begin{center}
\includegraphics[width=0.5\textwidth,angle=0]{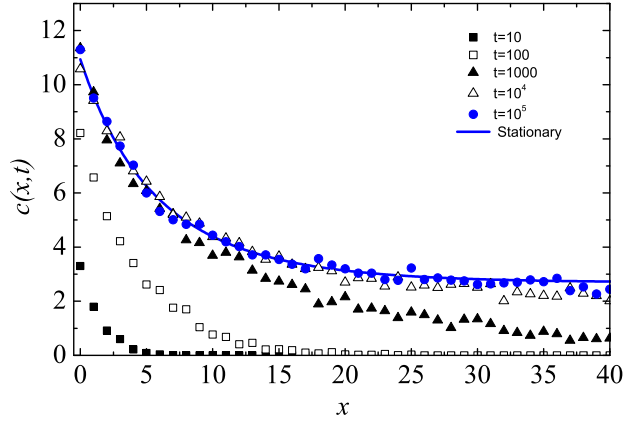}
\end{center}
\caption{\label{fig:normalProfile} Convergence of CTRW simulation results (symbols) to the stationary profile predicted by
formula (\ref{bessprof}) for $j_0=1,\gamma=1$, the corresponding value of the diffusion coefficient $K_\gamma=1/3$ and
exponentially decaying reactivity  $k(x)=k_0\exp(-\beta |x|)$ with $k_0=1/100$ and $\beta=1/8$ (solid line). CTRW jump
characteristics as in fig. \ref{fig:Escalon1}}
\end{figure}

\begin{figure}[ht]
\begin{center}
\includegraphics[width=0.5\textwidth,angle=0]{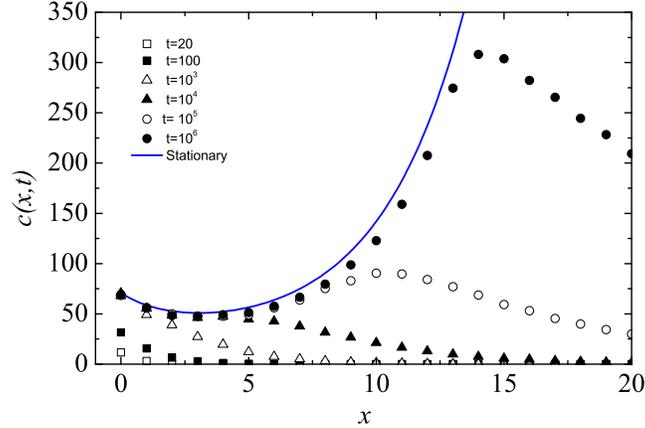}
\end{center}
\caption{\label{fig:subProfile} Convergence of CTRW simulation results (symbols) to the stationary profile predicted by the
formula (\ref{bessprof}) for  $j_0=1$, $\gamma=0.5$ , the corresponding value $K_\gamma=1/\sqrt{9\pi}$ and exponentially
decaying reactivity  $k(x)=k_0\exp(-\beta |x|)$ with $k_0=1/200$ and $\beta=0.6$ (solid line). CTRW jump characteristics as in
fig. \ref{fig:Escalon1}. The simulation results clearly go towards the stationary solution as time increases, although the
convergence for large $x$ is slow.} 
\end{figure}

(see Fig. \ref{fig:normalProfile}). In contrast, for $\gamma<1$ we find a qualitatively different behaviour. As one moves away
from the source, first the concentration decreases until it reaches a minimum and then it increases (see Fig.
\ref{fig:subProfile}).

\subsection{Power law reactivity}

Next, we take $k(x)= \kappa_0 (x_0+|x|)^{-\frac{\mu}{\gamma}}$ with $\mu>0$. Since a steady state was already attained for an
exponentially decaying reactivity, this will also be the case under the present situation, which describes enhanced particle
evanescence.The general solution of Eq. (\ref{vxunudelta}) for $\mu \neq 2$ is given by the modified Bessel functions
$I_{|\nu|}$ and $K_{|\nu|}$ with $\nu=(\mu-2)^{-1}$. In order to single out the Bessel function corresponding to the physical
solution, we use the fact that the incoming flux must be equal to the amount of particles per unit time that disappear due to
degradation, i.e.,

\be \label{fluxcond} j_0= \int_{-\infty}^\infty k(x) c_s(x) dx. \ee

This condition leads to different solutions depending on the value of $\mu$. For $\mu<2$ one gets

\be c_s(x)\propto (x_0+|x|)^{\frac{\mu}{\gamma}-\mu+\frac{1}{2}}K_{\left|\nu\right|}\left(\Phi
(x_0+|x|)^{-\frac{1}{2\nu}}\right), \ee

where $\Phi=2|\nu|\sqrt{\kappa_0^\gamma/K_\gamma}$. For large $|x|$, the above stationary solution can be shown to go to zero
as $x^{\frac{\mu}{\gamma}-\frac{3\mu}{4}}\,\exp{\left(-\Phi |x|^{1-\frac{\mu}{2}}\right)}$. In contrast, for $\mu>2$ one has

\be c_s(x) \propto (x_0+|x|)^{\frac{\mu}{\gamma}-\mu+\frac{1}{2}}I_{\left|\nu\right|}\left(\Phi
(x_0+|x|)^{-\frac{1}{2\nu}}\right). \ee As $|x|\to \infty$, this expression tends to a constant limiting value in the normal
diffusion case and grows as $x^{\frac{\mu}{\gamma}-\mu}$ for $\gamma<1$.

When $\mu=2$ the solution is not given by a power law rather than by Bessel functions. One has

\be c_s(x)\propto \left(x_0+x\right)^{\frac{2}{\gamma}-2+\lambda_-} \ee

with $\lambda_-=1\pm\sqrt{1+4(\kappa_0^\gamma/K_\gamma)}/2$. For $|x|\to\infty$ this solution goes to infinity, a constant
value or zero depending on whether $\frac{2}{\gamma}-2+\lambda_-$ is positive, zero or negative.

\section{Conclusions and Outlook}

In the present work we investigate both analytically and numerically the behaviour of the stationary concentration profiles
arising from fractional reaction-subdiffusion equations derived from a CTRW model with a superimposed death process. These
fractional equations are a natural extension of classical reaction-diffusion equations traditionally used to study the problem
of morphogen gradient formation. We consider the case of linear degradation with both a uniform and non-uniform reactivity.
The formulation of the problem in terms of fractional diffusion equations turns out to be a key ingredient in the analysis of
the properties of morphogen gradients. This approach allows us to exploit a plethora of powerful analytical techniques
available from fractional calculus to tackle the morphogen problem.

In the uniform case one obtains exponentially decaying stationary concentration profiles. Their robustness with respect to
changes in the incoming flux increases with increasing $k$. Likewise, one can study the robustness of the profiles with
respect to a perturbation in $k$ by introducing the quantity

\be {\cal R}_k= a \left(k \frac{\partial x}{\partial k}\right)^{-1} \ee

This issue will be the subject of future research.

In the non-uniform case the behaviour of the stationary profiles turns out to be very sensitive to the specific spatial
dependence prescribed for $k$ and to the value of the anomalous diffusion coefficient $\gamma$. Moreover, for the case of a
piecewise constant reactivity with $k_1=0$ and anomalous diffusion, we see that a discontinuous profile arises and no steady
state is reached in region $1$. This is a novel effect not seen for normal diffusion. For exponentially decaying reactivity
the concentration goes to a constant limiting value far away from the source ($|x|\to\infty$) when $\gamma=1$, but it grows
without bound for $\gamma<1$. Finally, when the connectivity decays as a power law, the stationary concentration may go to
zero, to a constant or to infinity depending on the values of the characteristic decay exponent and $\gamma$.

In view of the strong inhomogeneities encountered by the diffusing morphogens in the embryonic environment, we believe that
the sensitivity of the concentration gradients to the form of $k(x)$  may be relevant for the modeling of morphogen gradient
formation and interpretation.

Up to the case of piecewise constant reactivity with $k_1=0$, in the present work we limit ourselves to study the behaviour of
the stationary concentration profiles. However, analytic solutions for transient profiles are available for some of the cases
studied, and others can be investigated via numerical techniques for the inversion of Laplace transforms. Besides, a
reaction-subdiffusion equation was recently obtained that generalizes the above results to the general case $k=k(x,t)$
\cite{Yadav06, Yadav08}. As one could have guessed in view of Eqs. (\ref{unifreacsubdeq}) and (\ref{sekiEq}), this equation
reads as

\be \label{reacsubdi1} \frac{\partial c(x,t)}
 {\partial t}=    K_\gamma \frac{\partial^2 }{\partial x^2}
\left\{ e^{-\int_{0}^t k(x,t') dt'}   ~_{0}D_t^{1-\gamma}
 \left[ e^{ \int_{0}^t k(x,t') d t'}
 c(x,t)\right]\right\}-k(x,t)  c(x,t)
\ee

Our aim is to use the above equation to investigate further problems related to morphogen gradient formation in future. Beyond
this field of research, Eq. (\ref{reacsubdi1}) can be applied to many other problems of interest characterized by different
kinds of boundary conditions.

\acknowledgments This work was partially supported by the Ministerio de Ciencia y Tecnolog\'{\i}a (Spain) through Grant No.
FIS2007-60977, by the Junta de Extremadura (Spain) through Grant No. GRU09038, and by the National Science Foundation under
grant No. PHY-0855471.


\end{document}